\documentclass[twocolumn,showpacs,prX,nofootinbib,nobibnotes]{revtex4}

\usepackage{graphicx}
\usepackage{dcolumn}
\usepackage{bm}

\usepackage{anysize}
\usepackage{graphicx}
\marginsize{1.8 cm}{1.5 cm}{0.5 cm}{3.0 cm}

\usepackage{graphicx}
\usepackage{dcolumn}
\usepackage{bm}
\usepackage{amsmath}
\usepackage{amssymb}
\usepackage{mathrsfs}
\usepackage[latin1]{inputenc}
\usepackage{amsmath}

\newcommand{\dd}{{\rm d}}

\newcommand{\sd}{Schr\"{o}dinger }

\newcommand{\U}{\mathcal{U}}

\newcommand{\hil}{\mathcal{H}}

\newcommand{\tr}{{\rm Tr}}

\begin {document}


\title{Computational complexity of quantum optimal control landscapes}

\author{Raj Chakrabarti}
\affiliation{Department of Chemistry, Princeton University,
Princeton, New Jersey 08544, USA}
\email{rajchak@princeton.edu}

\author{Rebing Wu}
\affiliation{Department of Chemistry, Princeton University,
Princeton, New Jersey 08544, USA}

\author{Herschel Rabitz}
\affiliation{Department of Chemistry, Princeton University,
Princeton, New Jersey 08544, USA}

\date{23 August 2007}

\begin{abstract}

We study the Hamiltonian-independent contribution to the complexity of quantum optimal control problems. The optimization of controls that steer quantum systems to desired
objectives can itself be considered a classical dynamical system that executes
an analog computation. The system-independent component of the equations
of motion of this dynamical system can be integrated analytically for
various classes of discrete quantum control problems.
For the maximization of observable expectation values from an initial pure state and the
maximization of the fidelity of quantum gates, the time complexity
of the corresponding computation belongs to the class continuous log (CLOG), the
lowest analog complexity class, equivalent to the discrete
complexity class NC. The simple scaling of the Hamiltonian-independent contribution to these problems with quantum system dimension indicates that with appropriately designed
search algorithms, quantum optimal control can be rendered efficient
even for large systems.

\end{abstract}


\maketitle

\section{Introduction}

In recent years the methodology of optimal control theory has been
applied to achieve various objectives in quantum systems. The two
most common objectives are the maximization of the expectation value
of an observable, and the maximization of the fidelity of a unitary
transformation of the wavefunction. Whereas the maximization of
observable expectation values has met with widespread success in
experimental and computational incarnations \cite{ShaBru2006}, the
achievement of high fidelity unitary transformations has proven more
challenging \cite{Kosloff2002}, especially for large systems.
However, it is not clear whether these apparent distinctions are
specific to the algorithms employed, or whether they are indicative
of universal and inherent features of the quantum optimization
problems.

Given the computational expense of propagating the
\sd equation, the scaling with system size of the expense of
executing a search over the space of control fields -- referred to as the problem's computational complexity -- plays an essential role in determining the feasibility and efficiency of
control optimizations in large quantum systems. The complexity of control optimization
for quantum gates is of particular interest, as it corresponds to the classical complexity associated with the physical implementation of quantum logical operations. Recent studies have addressed the
computational complexity (and related, the critical topology) of several associated problems in quantum information science, including quantum state and process identification \cite{Lidar2007}.
In contrast to these tasks, the variational problem of quantum control optimization occurs on an infinite-dimensional parameter space. However, for discrete quantum systems, both the complexity and critical topology of quantum control problems can be divided into finite- and infinite-dimensional
contributions, simplifying their study.

Consider first the critical topology of quantum control variational problems.
Denoting the cost functional by $J$ and the control field by $\varepsilon(t)$, we have according to the
chain rule $$\frac{\delta J}{\delta \varepsilon(t)} = \frac{\dd J}{\dd U(T)} \cdot
\frac{\delta U(T)}{\delta \varepsilon(t)},$$ where $U(T)$ is the dynamical propagator resulting from application of the control $\varepsilon(t)$ over the time interval $0 \leq t \leq T$. As such, the critical points of quantum control problems
fall into two classes. The first type of minimizer corresponds to those control Hamiltonians that are
critical points of the control objective functional, but are not
critical points of the map between control fields and associated
dynamical propagators (i.e., points at which $\frac{\dd J}{\dd
U(T)}=0$, while the Frechet derivative mapping from the control
variation $\delta \varepsilon(t)$ to $\delta U(T)$ at $t=T$ is
surjective). These critical points are called normal extremal controls.
The second type of minimizer corresponds to critical points of the latter map
(i.e., points at which the mapping from $\delta \varepsilon(t)$ to
$\delta U(T)$ is not locally surjective). These latter points are referred to
as abnormal extremal controls. For several low-dimensional quantum control problems,
it has been shown that abnormal extremals are particularly scarce \cite{Raj2007,Wu2007,DAlessandro2001a}, suggesting that
the normal extremal controls dominate the critical point topology of these problems.

As such, attention has focused on characterizing the system-independent (or kinematic) contribution to the critical topology of quantum optimal control problems \cite{RabMik2004,Mike2006a,Mike2006b}. For a given quantum system, several different classes of quantum control problems may be envisioned, each of which has its own associated kinematic critical topology. Besides the distinction between quantum gate and observable control, observable control itself may be subdivided into several qualitatively distinct optimization
problems, depending on the initial state (density matrix) of the system and the type of observable
to be maximized. Mathematically, the critical topologies of observable control problems can be classified
on the basis of the eigenvalue spectra of the initial density matrix $\rho_0$ and the observable
operator $\Theta$ \cite{Mike2006a}. To date, however, the system-independent contribution to the computational complexity of these problems has not been addressed.  Here, we show that for several classes of quantum control problems, this complexity can, remarkably, be analytically determined.

\section{Control optimization as a dynamical system}

A universal quantum optimal control cost functional can be written
as:
\begin{multline}
J =\Phi(U(T), T)-\\
\textmd{Re} \tr\left[\int_{0}^T\left\{\left(\frac{\partial U(t)}{\partial t} +
\frac{i}{\hbar}H(\varepsilon(t))U(t)\right)B(t)\right\}dt\right]\\
-\lambda\int_{0}^T\frac{1}{s(t)}|\varepsilon(t)|^2 dt
\end{multline}
where $B(t)$ is a Lagrange multiplier constraining the quantum
system dynamics to obey the \sd equation, $\varepsilon(t)$ denotes the
time-dependent control field, $s(t)$ denotes the pulse envelope,
and $\lambda$ weights the importance of the penalty on the total
field fluence. Solutions to the optimal control problem correspond
to $\frac{\delta J}{\delta \varepsilon(t)} = 0$. The problem of maximizing
the expectation value of an observable corresponds to:
\begin{equation}
\Phi_1(U) = \tr(U(T){\rho_0}U^{\dag}(T)\Theta)
\end{equation}
where $\rho_0$ is the initial density matrix of the
system and $\Theta$ is an arbitrary observable operator, and the
problem of maximizing the fidelity of a dynamical transformation $W$
then corresponds to \cite{Kosloff2002}:
\begin{equation}
\Phi_2(U) = Re~\tr(AW^{\dag}U(T))
\end{equation}
where $W$ is the target unitary transformation and A is any
Hermitian matrix. In the present study, we ignore the fluence
penalty term because its effect is system-specific and not revealing
with respect to the universal complexity of quantum control
problems.

We are interested in the convergence to the global optimum of the
gradient flow trajectories of the objective functions, which are the
solutions to the differential equations
\begin{equation}\label{Egrad}
\frac{d\varepsilon(s,t)}{ds}= \triangledown \Phi(\varepsilon(t)) =\alpha \frac
{\delta  \Phi_{1,2}(s,T) }{\delta \varepsilon(s,t)}
\end{equation}
where $s$ is a continuous variable parametrizing the algorithmic time evolution of
the search trajectory and $\alpha$ is an
arbitrary scalar that we will set to 1. The complexity of quantum control optimization is associated with the
scaling with system size of the rate of convergence of the function
$$ \Phi_{1,2}(s,T)-\Phi_{1,2}(\infty,T) = \alpha \int_{\infty}^s ds'
 \int_0^T dt \left[\frac {\Phi_{1,2}(s',T)}{\delta \varepsilon(s',t)}\right]^2
 $$ to zero. The gradients  $\frac{\delta \Phi}{\delta \varepsilon}$ can be shown
to be, respectively \cite{HoRab2007a,HoRab2007b,RajWu2007}:

\begin{equation}\label{grad1}
\frac{\delta \Phi_1}{\delta \varepsilon(t)} =\tr
\big([\Theta(T),\rho_0]\mu(t)\big),
\end{equation}
and \begin{equation}\label{grad2} \frac{\delta \Phi_2}{\delta
\varepsilon(t)} =\tr \left((AA^{\dag}W^{\dag}U-U^{\dag}WAA^{\dag})\mu(t)\right),
\end{equation}
within the electric dipole approximation, where $\mu(t) = -\frac{i}{\hbar}U^{\dag}(t,0)\mu U(t,0)$ is the time-evolved dipole operator of the quantum system and we have adopted the shorthand notation $U \equiv U(T)$.

The gradient on $\varepsilon(t)$ is related to the gradient on $\U(N)$ through
\begin{equation}\label{chain}%
\frac{\delta \Phi}{\delta \varepsilon(t)}=\sum_{i,j}\frac{\delta
U_{ij}}{\delta \varepsilon(t)}\frac{\dd \Phi}{\dd U_{ij}}.
\end{equation}
Now suppose that we have the gradient flow of $\varepsilon(s,t)$
that follows (\ref{Egrad}) and let $U(s)$, the system propagator at
time $T$ driven by $\varepsilon(s,t)$, be the projected trajectory
on the unitary group $\U(N)$. The (algorithmic) time derivative of
$U(s)$ is then
\begin{equation}\label{Us}
  \frac{\dd U_{ij}(s)}{\dd s}=  \int_0^T \frac{\delta U_{ij}(s)}{\delta \varepsilon(s,t)}\frac{\partial
\varepsilon(s,t)}{\partial s} \dd t
\end{equation}
which, combined with (\ref{Egrad}) and (\ref{chain}), gives
\begin{equation}\label{dot Us}
     \frac{\dd U_{ij}(s)}{\dd s}=\int_0^T \frac{\delta U_{ij}(s)}{\delta \varepsilon(s,t)}\sum_{p,q}\frac{\delta U_{pq}(s)}{\delta \varepsilon(s,t)}\frac{\dd \Phi}{\dd U_{pq}} \dd t.
\end{equation}
It is convenient to write this equation in vector form, replacing
the $N \times N$ matrix $U(s)$ with the $N^2$ dimensional vector
$\textbf{u}(s)$:
\begin{multline}\label{vecu}
\frac{\dd \textbf{u}(s)}{\dd s}=\left[\int_0^T
\frac{\delta\textbf{u}(s)}{\delta
\varepsilon(s,t)}\frac{\delta\textbf{u}^T(s)}{\delta
\varepsilon(s,t)}\dd t\right]\triangledown
\Phi\left[\textbf{u}(s)\right] \\
\equiv \textmd{G}[\varepsilon(s,t)]\triangledown \Phi\left[\textbf{u}(s)\right]
\end{multline}
where the superscript $T$ denotes the transpose. This
relation implies that the variation of the propagator in $\U(N)$
caused by the natural gradient flow in the space of control fields is
Hamiltonian-dependent, where the influence of the Hamiltonian is
contained in the $N^2$-dimensional symmetric matrix
$\textmd{G}[\varepsilon(s,t)]$. Within the electric dipole approximation, $\textmd G$
is given by
$$\textmd{G}_{ij,pq}(s) = \int_0^T \mu_{ij}(s,t)\mu_{pq}(s,t) \dd t.$$

As we will discuss below, it is possible to eliminate $\textmd{G}$ in this differential equation,
and hence to explicitly follow the $U(T)$ gradient flow, by adopting an alternative algorithmic step on $\varepsilon(t)$. Importantly, the properties of the map $\varepsilon(t) \rightarrow U(T)$ for finite-dimensional quantum systems render this fairly simple to achieve across many diverse systems, albeit with an error that is Hamiltonian-dependent. For the present purpose, we can effectively ignore $
\textmd{G}$ because we are interested in comparing the complexity scalings of the flow trajectories for gate and observable control problems on the same quantum system. Averaged over many initial conditions,
the time required for convergence of the $U(T)$ gradient flows will determine this comparative scaling.

Because the kinematic gradient flows associated with these differential equations evolve on the
continuous space of a Lie group (for $\Phi_2$) or its adjoint
orbit of skew-Hermitian matrices (for $\Phi_1$), the optimization
processes can be considered to be analog rather than discrete
computations. In prior work \cite{RajWu2007}, we demonstrated that
these flow trajectories can in fact be viewed as dynamical systems whose equations of motion
guide the algorithmic time evolution of gradient-based algorithms. We showed that although the infinite-dimensional gradient flow
equations do not admit analytical solutions, the equations of motion
for the gradient flow lines on the unitary group domain, namely
\begin{eqnarray*}
  \left(\frac{dU}{ds}\right)_1 &=& -U[\rho_0,U^{\dag}\Theta
U~], \\
  \left(\frac{dU}{ds}\right)_2 &=& A^{\dag} - UAU
\end{eqnarray*}
can in principle be exactly solved and explicitly integrated them
for specific cases of interest. The solutions to the optimal control problem are the critical points of the flow, i.e. $\triangledown
\Phi(U) = 0$, or equivalently, the equilibria of the dynamical
system. In this work, we use the integrated gradient flow equations to derive
upper bounds on the time for convergence to the solutions of these
problems and thereby assign the problems to computational
complexity classes.

The possible outputs of the dynamical system executing the analog
computation are the attracting fixed points of the gradient flow of
the control objective functional. Just as a discrete computation can
be assigned a measure of time complexity, i.e. the scaling with
system size (e.g., polynomial or exponential) of the time required
to solve the problem, analog computations can be assigned time
complexities, which correspond to the scaling with system size of
the time required for the dynamical system to converge to a vicinity
of the optimal fixed point. Time complexity of an analog computation
can be analytically determined for systems that are completely
integrable; i.e., for those systems whose equations of motion can be
exactly solved. For this purpose, we apply the recently developed
theory of complexity for continuous time dynamics
\cite{Fishman1999,Fishman2002}.

For $\Phi_1$, the gradient flow evolves on a polytope whose
dimension varies depending on the spectrum of $\rho_0$. Since the
flow is cubic in $U$, many trajectories $U_1(s)$ correspond to the
same function $J(s)$. Rewriting the gradient flow as a quadratic
function on the domain of Hermitian operators $\rho_T=U(T)\rho_0
U^{\dag}(T)$ \cite{RajWu2007}, one can solve explicitly for a unique solution
$\rho_T(s)$ corresponding to $J(s)$. In the case that $\rho_0$ has
only one nonzero eigenvalue, corresponding to an initial pure state,
we showed that under the change of variables $\rho_T(s) =
|\psi(s)\rangle \langle \psi(s)\rangle $, $|\psi(s)\rangle  =
(c_1(s),\cdots,c_N(s))$, $x(s) \equiv
(|c_1(s)|^2,\cdots,|c_N(s)|^2)$, the gradient flow can be explicitly
integrated to give \cite{RajWu2007}:
\begin{eqnarray}
x(s) &=& \frac{e^{2s\Theta}\cdot(|c_1(0)|^2,\cdots,|c_N(0)|^2)}{\sum_{i=1}^N|c_i(0)|^2e^{2s\lambda_i}} \\
&=&\frac{e^{2s\lambda_1}|c_1(0)|^2,\cdots,e^{2s\lambda_N}|c_N(0)|^2}{\sum_{i=1}^N|c_i(0)|^2e^{2s\lambda_i}}
\end{eqnarray}
where $\lambda_1,\cdots, \lambda_N$ denote the eigenvalues of
$\Theta$. The explicit solution for the gradient trajectory of
objective functional $\Phi_2$ was shown to be
\begin{multline}
W^{\dag}U(s) = (\sinh (As) + \cosh (As) W^{\dag}U_0)\cdot \\
\dot(\cosh(As) + \sinh (As) W^{\dag}U_0)^{-1}
\end{multline}
where the initial condition is $U_0 = U(0)$ \cite{RajWu2007}.

In the case of objective functional $\Phi_1$, the input for the
computation may be viewed as being the eigenvalues of the observable
operator, while the initial density matrix determines the initial
condition. For the purpose of assigning a complexity class to the
problem, the latter is generally taken as fixed. The time complexity
is then expressed as a function of the eigenvalues of $\Theta$.
However, as discussed below, the complexity of the optimal control
problem can vary sharply as a function of the initial state as well.
For objective functional $\Phi_2$, we consider the input for the
computation to be the target unitary transformation $W$. The time
complexity is expressed in terms of the eigenvalues of $W$. Again,
time complexity can be modulated by varying the initial condition of
the search away from the identity matrix $I$.

It is possible to assess the convergence of the optimal control
search in terms of the difference in objective function values
$\Phi(s) - \Phi(\infty)$, but because of the degeneracy of solutions
corresponding to a given value of $\Phi$, a more precise assessment
of complexity can be achieved in terms of the convergence of the
distance between the current guess and the global solution to the
problem. As mentioned, the distance $||E(s)-E(\infty)||^2 = \int_0^T
E(s, t) E(\infty, t) dt$ is not an appropriate measure because it is
highly system-specific (precluding uniform behavior as a function of
system size) and obscures the underlying geometry of convergence
since $\Phi$ is not explicitly a function of $E(t)$. The most
appropriate choice is a distance on the space of solutions that
displays the lowest degeneracy for a given value of $\Phi$, while
remaining independent of the system Hamiltonian. For $\Phi_1$, this
distance is $||U(s)-U(\infty)||_F^2$, whereas for $\Phi_2$, it is
the Euclidean distance on the polytope wherein the flow evolves.

The attracting region $R$ of an attracting fixed point is the subset
of phase space wherein the distance to the point is monotonically
decreasing with time. We consider the quantum control optimization
problem to be solved to a desired precision, and the computation
halted, if the gradient flow trajectory enters within an
$\epsilon_p$-radius of the global optimum and is also within its
attracting region.  This definition ensures that for inputs on which
the $\epsilon_p$-vicinity of the attractor is larger than the
attracting region, entrance into the attracting region is required
for halting of the computation. The computational complexity of the
control optimization algorithm is then defined as the scaling with
system size of the convergence time
$t_c(H)=\max\left[t_c(\epsilon),t_c(R)\right]$ to the intersection of the
attracting region and the ball of radius $\epsilon_p$, starting from
a given initial condition.

\section{Analog complexity of quantum control optimization}

\subsection{Quantum observable maximization}

For optimization of $\Phi_1$ starting from an initial pure state,
we can establish a bound on $t_c(\epsilon)$ as follows. We assume
without loss of generality that the eigenvalues of $\Theta$ are
arranged such that $\lambda_1 \geq \lambda_2 \geq \lambda_3
\geq\cdots\geq \lambda_N$. For nondegenerate $\Theta$, the global
optimum is then the unit vector $e_{i*}=e_{1}$ and the distance of
the current point $x(s)$ on the search trajectory to the global
optimum can be written:
\begin{multline}
||x(s)-x(\infty)||^2 =\\
||x(s)||^2-2\frac{(e^{2s\lambda_1}|c_1(0)|^2,\cdots,e^{2s\lambda_N}|c_N(0)|^2)\cdot
e_1}{\sum_{i=1}^N|c_i(0)|^2e^{2s\lambda_i}}+1\\
\leq2-2\frac{e^{2s\lambda_{1}}|c_{1}(0)|^2}{\sum_{i=1}^Ne^{2s\lambda_i}|c_i(0)|^2}
\end{multline}
since $\sum_{i=1}^N x_i = \sum_{i=1}^N c_i^2 = 1$ and therefore
$\sum_{i=1}^N x_i^2 = \sum_{i=1}^N c_i^4 \leq 1$. Defining $\mu
\equiv \lambda_1-\lambda_2$, we then have
$$\sum_{i=1}^N e^{2s(\lambda_i-\lambda_{1})}|c_i(0)|^2 \leq e^{-2 \mu s} +
|c_{1}(0)|^2$$ such that the bound on the distance to the solution
becomes
$$||x(s)-x(\infty)||^2 \leq 2-2(1+e^{-2\mu s}|c_{1}(0)|^{-2})^{-1}.$$
For simplicity, we choose the identity vector $\frac{1}{N}(1,...,1)$
as the initial state. More generally, $x_{i*}(0)$ will scale
inversely with $N$ for randomly chosen initial conditions on the
interior of the simplex. We then obtain the following bound on the
convergence time to the global optimum:
$$t_c(\epsilon) \leq \Big|\frac{\ln(\frac{\epsilon^2 |c_{1}(0)|^2}{2})}{2\mu}\Big| = \frac{1}{2\mu}\ln\Big(\frac{2N}{\epsilon^2}\Big)$$
for $\epsilon$ small. Note that the time scale for convergence to
the global optimum (the linearization of the coefficient in the
exponential) is given by $\mu$. This is equal to the lowest
eigenvalue of the Hessian of the objective function,
$\hil_A(\rho(\infty))=-\sum_{j\neq
1}(\alpha_{j1}^2+\beta_{j1}^2)(\lambda_j-\lambda_1)$, where
$\alpha_{jk}, \beta_{jk}$ are the real, complex parts of an
arbitrary Hermitian matrix $A$, obtained in \cite{Mike2006a}
\footnote{It can be shown that the expression derived in
\cite{Mike2006a} for $\hil_A(\U(\infty))$ is equivalent to that for
$\hil_A(\rho(\infty)).$}. If the maximum eigenvalue of $\Theta$ is
degenerate with multiplicity $k$, such that
$\lambda_1=\lambda_2=\cdots=\lambda_k$, and $\lambda_k > \lambda_j,
\quad j=k+1,\cdots,N$, then the dynamics converges to the point
$\frac{1}{k}(1,\cdots,1,0,\cdots,0)$ \cite{RajWu2007}. In this case,
the distance to the global optimum becomes
\begin{multline}
||x(s)-\frac{1}{k}(1,\cdots,1,0,\cdots,0)||^2 \leq \\
2-2\frac{\frac{e^{2s\lambda_1}}{k}\sum_{i=1}^k
|c_i(0)|^2}{\sum_{j=1}^N e^{2s\lambda_j}|c_j(0)|^2},
\end{multline}
corresponding to $$t_c(\epsilon) \leq
\frac{1}{2\mu}\ln\left(\frac{2Nk}{\epsilon^2}\right).$$

A bound on the convergence time to the attracting region of the
solution, $t_c(R)$, can be derived as follows for the general case
of a $k$-fold degenerate maximal eigenvalue of $\Theta$. Again,
without loss of generality, we assume $\lambda_{1,...,k} \equiv
\lambda_{(1)} > \lambda_{k+1}>\cdots>\lambda_N$. In the present
case, $t_c(R) = t_c$ such that for all $s
> t_c$,
\begin{multline}
\quad \frac{\partial}{\partial s} ||x(s)-x(\infty)||^2 =\\
e^{2s\lambda_1}\sum_{i=1}^k|c_i(0)|^2\sum_{j=1}^N~[(\lambda_{(1)}-\lambda_j)e^{2s\lambda_j}|c_j(0)|^2~]
< 0.
\end{multline}
This corresponds to the condition $\dot x_i < 0$ for $i > k$
\cite{RajWu2007}. This condition holds iff $\sum_{j=1}^N
\lambda_jx_j
> \lambda_i, \quad i=k+1,\cdots,N$. Insertion of the analytical
solution above gives
$$\sum_{j=1}^N \lambda_j|c_j(0)|^2e^{\lambda_j s} >  \lambda_{k+1} \sum_{j=1}^N
|c_j(0)|^2 e^{\lambda_js}.$$ For the purposes of obtaining a bound
on $t_c$, we can rewrite this
\begin{multline}
\sum_{i=1}^k(\lambda_{(1)}|c_i(0)|^2-\lambda_{k+1}|c_{k+1}(0)|^2)e^{\lambda_{(1)}t_c} >\\
(N-k-2)\lambda_{k+1}|c_{k+1}(0)|^2e^{\lambda_{k+1}t_c},
\end{multline}
which can be solved for $t_c(R)$ to give:
\begin{multline}
t_{c,1}(R) \leq \frac{1}{\lambda_{(1)}-\lambda_{k+1}}\cdot\\
\cdot \ln\left\{
\frac{(N-k-2)\lambda_{k+1}|c_{k+1}(0)|^2}{\sum_{i=1}^k~[\lambda_{(1)}|c_i(0)|^2-\lambda_{k+1}|c_{k+1}(0)|^2~]}\right\}.
\end{multline}
Therefore, the attracting region of the solution
$x(\infty)=\frac{1}{k}(1,\cdots,1,0,\cdots,0)$ does not cover the
entire domain. Again taking the initial condition
$\frac{1}{N}(1,...,1)$, this bound becomes $t_{c,1}(R) \leq
\frac{1}{\mu}\ln\frac{(N-k-2)\lambda_{k+1}}{k(\lambda_{1}-\lambda_{k+1})}.$
The upper bound on the computation time of the problem is then
given by
\begin{multline}
t_{c,1}(H) = \textmd{max}~[t_{c,1}(\epsilon),t_{c,1}(R)~]\leq\\
\frac{1}{2\mu}\Big[\ln\Big(\frac{2Nk}{\epsilon^2}\Big)+2\ln\frac{(N-k-2)\lambda_{k+1}}{k(\lambda_{(1)}-\lambda_{k+1})}\Big].
\end{multline}

\subsection{Quantum gate control}
For objective function $\Phi_2$, the control optimization is
generally initiated with $U_0$ at or near the identity
transformation $I_N$, such that $W^{\dag}U_0 = W^{\dag}.$ Writing
$U'(s)\equiv W^{\dag}U(s)$, the solution to the problem then
corresponds to $U'(\infty)= I_N$. We consider the special case where
the Hermitian matrix $A=I$:
$$U'(s)=(\sinh (sI) + \cosh (sI)W^{\dag})(\cosh (sI) + \sinh (sI)W^{\dag})^{-1}$$
For $\Phi_2$, the natural distance measure for assessment of
convergence is given by the Frobenius matrix norm between the
current transformation $U'(s)$ and the solution $I_N$:
$$||U'(s)-U'(\infty)||_F^2 = \tr [(U'(s)-I_N)^{\dag}(U'(s)-I_N)]$$
Diagonalizing $W$ via the unitary transformation $V$, we have
$$W = V^{\dag} \Lambda V$$
where $\Lambda$ is the matrix of eigenvalues of $W$,
\begin{equation}
\Lambda=\left(%
\begin{array}{ccc}
  e^{-i\theta_1} &  &  \\
   & \ddots &  \\
   &  & e^{-i\theta_N} \\
\end{array}%
\right)%
\end{equation}
We previously showed \cite{RajWu2007} that this transformation
allows the expression above to be simplified as
\begin{multline}
||U'(s)-I_N||^2 = \sum_{k=1}^N \Big|\frac{\tanh s +
e^{-i\theta_k}}{1+ e^{-i\theta_k}\tanh s}-1 \Big |^2=\\
\sum_{k=1}^N \frac{2(1-\cos\theta_k)}{1+2\tanh s \cos \theta_k +
\tanh^2 s}(1-\tanh s)^2
\end{multline}
The partial derivative of each term in this expression with respect
to $\cos\theta_k$ monotonically decreases with $\cos\theta_k$, so
the greatest term in the sum will correspond to the lowest value of
$\cos\theta_k$, denoted $\cos\theta_0$. Therefore, the sum is
bounded from above by $N$-times the term corresponding to
$\theta_0$,
$$||U'(s)-I_N||^2 \leq N \frac{2(1-\cos\theta_0)}{1+2\tanh s \cos \theta_0 +
\tanh^2 s}(1-\tanh s)^2$$ Under the change of variables $x \equiv
\tanh s$, the condition for entering an $\epsilon$-vicinity of the
solution is
$$N \frac{2(1-\cos\theta_0)}{1+2x \cos \theta_0 + x^2}(1-x)^2 <
\epsilon.$$  To derive tight upper bounds on $x_{c}$ and hence
$t_{c}$, we must therefore solve the quadratic equation $(2N-2N \cos
\theta_0 - \epsilon)x_{c,\textmd{max}}^2 +
(4N\cos\theta_0-4N-2\epsilon \cos \theta_0)x_{c,\textmd{max}} +
(2N-2N\cos\theta_0-\epsilon) = 0$. The solutions to this equation
are:
\begin{widetext}
$$x_{c,\textmd{max}} \equiv \tanh t_{c,\textmd{max}} = \frac {4N-4N\cos\theta_0 + 2\epsilon \cos \theta_0 \pm
\sqrt{(\epsilon^2-4N\epsilon)(\cos^2\theta_0-1)}}{4N-4N\cos\theta_0-2\epsilon}.$$
\end{widetext}
One solution gives identically $x_{c,\textmd{max},+} = 1$,
corresponding to $ t_{c,\textmd{max},+} = \infty$, which only
provides an upper bound on the convergence time. The other solution
gives $\quad x_{c,\textmd{max},-} \leq 1, \quad
t_{c,\textmd{max},-} \leq \infty$. In order to study the scaling of
$t_{c,\textmd{max},-}$ with the dimension of the system, we consider
the deviation of $x_{c,\textmd{max},-}$ from 1:
\begin{multline}
\delta \equiv 1-x_{c,\textmd{max},-} =\\
\frac {2\epsilon(1-\cos\theta_0) + \sqrt{(\epsilon^2-4N\epsilon)(\cos^2\theta_0-1)}}{4N(1-\cos\theta_0)+2\epsilon}
\approx \\ \frac
{\sin\theta_0}{1-\cos\theta_0}\sqrt{\frac{\epsilon}{N}}=
a\sqrt{\frac{\epsilon}{N}}
\end{multline}

for small $\epsilon$. This corresponds to
\begin{multline}
t_{c}(\epsilon) \leq t_{c,\textmd{max},-}(\epsilon) = \ln \Big(\frac{1+x_{c,\textmd{max},-}}{1-x_{c,\textmd{max},-}}\Big) = \\
\ln \Big(\frac{2-\delta}{\delta}\Big) \approx \ln \Big(\frac{2-a\sqrt{\frac{\epsilon}{N}}}{a\sqrt{\frac{\epsilon}{N}}}\Big)
= \frac{1}{2}\ln \Big(\frac{4N}{a^2\epsilon}\Big)
\end{multline}
For $\Phi_2$, the distance to the solution $U'(T) = I_N$, $U(T) =
W^{\dag}$ is a monotonically decreasing function of algorithmic time
$s$ \cite{RajWu2007}, and hence in this case, the attracting region
of the solution covers the entire manifold $U(N)$ and $t_c(R)=0$.
Thus, an upper bound on the computation time for the problem is
\begin{equation}
t_{c,2}(H) = t_{c,2}(\epsilon) \leq \frac{1}{2}\ln
\Big(\frac{4N}{a^2\epsilon}\Big)
\end{equation}
for small $\epsilon$.

\subsection{Assignment to complexity classes}

A continuous time problem is said to be in the complexity class CLOG
(continuous log) if it has a polynomial number of variables (here,
the system dimension) and a logarithmic time complexity.
Kinematic optimization of objective functions $\Phi_1$ and $\Phi_2$ therefore
belong to the complexity class CLOG. CLOG is the analog counterpart
of the classical complexity class $\textmd{NC}_1$, the class of
problems that can be efficiently solved on a discrete parallel
computer, meaning problems that are decidable in polylogarithmic
time on a parallel computer with a polynomial number of processors.
It is the lowest time complexity class, lying immediately below P
(polynomial time complexity). For real control optimization
algorithms evolving in discrete time, the kinematic component of these problems will have
$\textmd{NC}_1$ complexity.

Quantum optimal control problems aimed at maximizing the expectation
value of any observable operator starting from a pure state of the
system all belong to the same analog time complexity class. However,
their characteristic time scale for exponential convergence, when
the optimization algorithm follows the gradient flow trajectories as
faithfully as possible, differs in a predictable fashion.
Specifically, the characteristic time scale for exponential
convergence is given by $\min_{j\neq 1}|\lambda_j - \lambda_1|$,
i.e., the magnitude of the difference between the largest and second
largest eigenvalue of the observable operator. This corresponds to
the magnitude of the smallest Hessian eigenvalue of the objective
functional near the solution. As such, the rate of convergence for
the problem of driving a pure initial state to a pure final state -
which corresponds to $\lambda_1=1, \quad \lambda_i=0 \quad \forall
i>1$ above - is the greatest, such that this problem has the lowest
computational complexity of all observable maximization problems.
The complexity classes corresponding to observable maximization
problems starting from general mixed states may differ and are
expected to be higher, as evidenced by 1) the factorial (vs
exponential) scaling of the number of critical manifolds
\cite{Mike2006a} and 2) the fact that the gradient flow evolves on a
higher dimensional polytope \cite{HoRab2007b,RajWu2007}.

Similarly, the kinematic component of the problem of optimizing controls for
the implementation of a quantum gate belongs to the complexity class CLOG irrespective
of the identity of the target gate $W$. In contrast to observable
expectation value maximization, however, the characteristic time
scale for exponential convergence to the solution of quantum gate
optimization does not vary directly as a function of the particular
incarnation of the problem embodied in the choice of $W$, since the
eigenvalues of the Hessian matrix do not vary as a function of the
eigenvalues of $W$ \cite{Mike2006b}. Nonetheless, the eigenvalue
spectrum of $W$ does affect the convergence time indirectly in
conjunction with the initial guess $U_0$.

Although the problems belong to the same complexity class with
respect to the scaling of the number of iterations required as a
function of system size, they display distinct behavior in other
regards. In particular, it can be shown \cite{RajWu2007} that for
both problems, there exists a set of initial conditions from which
the flow does not converge to the solution. For $\Phi_2$, this
corresponds to the case where $\cos~ \theta_0 = -1$, resulting in
$x_{c,-}=1$,  $t_{c,-} = \infty$. Thus, if $W$ has at least one
eigenvalue equal to $-1$, the optimization does not converge if
$U_0$ is chosen as $I$, since $U'(0) = W^{\dag}U_0=W^{\dag}$. For
$\Phi_1$, initial conditions where $c_i*=0$ (i.e., where the initial
state resides on the so-called basin boundary of the simplex) do not
converge. The scaling of the convergence time as a function of the
distance to this pathological initial condition differs for the two
problems \cite{RajWu2007}. In these cases, the initial guess $U_0$
should be modulated to facilitate the convergence of gradient
algorithms.

The complexity of optimal control landscapes for the optimization of
quantum gates scales linearly with the number of qubits. If the
problem size is measured in qubits, optimization of $\Phi_2$ belongs
to the complexity class CP (continuous P), the set of problems with
a polynomial number of variables and polynomial time complexity on a
continuous variable computer. CP is the analog counterpart of the
discrete complexity class P. In previous work aimed at optimizing
quantum gates through OCT, the scaling of the number of control
field iterations was reported to be exponential in the number of
qubits \cite{Kosloff2003}. In these studies, the gradient flow was
not followed directly by the optimization algorithm. The results
here suggest that exponential speedup should in principle be
possible for the implementation of quantum gates through the use of
algorithms that follow the gradient flow on the unitary group.

As such, a central issue concerns the accessibility of these lower
bounds on scaling through the design of gradient-based algorithms in
both the laboratory and in simulations. These continuous time
complexities represent lower bounds on the complexities actually
achieved by discretized Euclidean search algorithms that do not
precisely follow the Riemannian curvature of the gradient flow
trajectories. The increase in complexity induced by (Euclidean)
discretization depends on the curvature of the trajectory. A
quantitative analysis is beyond the scope of the current work, but
the path length of the gradient flow trajectory offers insight into
its geometry. On the unitary group $\U(N)$, the path length of the
gradient flow trajectory is given by
$$L(U_0,t_c)=\int_0^{t_c}\sqrt{\tr(\dot U^\dag(s)\dot U(s))}\dd s,$$
where the Riemannian metric on the unitary group is defined as
$\langle A,B\rangle=\tr(A^\dag B)$. For the optimization $\Phi_2$,
this integral can be shown to be
$$L(U_0,t_c)=2\int_0^{t_c}\sqrt{N-Re\tr~[(W^\dag U)^2~]} \textmd{ds},$$
which can be expressed as a sum of $N$ scalars:
\begin{widetext}
$$L(U_0,t_c)=2\int_0^{t_c} \textmd{ds}\sqrt{\sum_{k=1}^N \frac{2\sin^2\theta_k(1-\tanh ^2s)^2}{(1+\tanh^2s)^2+4\tanh s~[\cos^2\theta_k+(1+\tanh^2s)\cos\theta_k~]}}.$$
\end{widetext}
Again setting $x=\tanh t$, the integrand can be further simplified
using the relation
\begin{multline}
\sqrt{\frac{2N\sin^2\theta_k(1-x^2)^2}{(1+x^2)^2+4x~[\cos^2\theta_k+(1+x^2)\cos\theta_k~]}}\leq\\
\sqrt{\frac{2N\sin^2\theta_k(1-x^2)^2}{(1+x^2)^2(1-x)}}. \end{multline}
Then we have
\begin{eqnarray*}
  L(U_0,t_c) &\leq& \sqrt{2N}|\sin\theta|\int_{0}^{x_c}\frac{(1-x^2)}{(1+x^2)\sqrt{1-x}}\dd \ln \left(\frac{1+x}{1-x} \right)\\
   &=&  2\sqrt{2N}|\sin\theta|\int_{0}^{x_c}\frac{\dd x}{(1+x^2)\sqrt{1-x}} \\
   &<&   2\sqrt{2N}\frac{|\sin\theta|}{\sqrt{1-x_c}}\int_{0}^{x_c}\frac{\dd x}{1+x^2}  \\
   &=&  2\sqrt{2N}\frac{|\sin\theta|\arctan x_c}{\sqrt{1-x_c}}\\
   &<&  \pi\sqrt{2N}\frac{|\sin\theta|}{\sqrt{1-x_c}}\sim |\sin\theta| \epsilon^{-1/4}N^{3/4}.\\
\end{eqnarray*}Therefore, we
see that the path length - which unlike the convergence time, does
not depend directly on the magnitude of the gradient and hence is
less sensitive to the effects of discretization - scales
approximately linearly with $N$, roughly consistent with the scaling
observed in kinematic simulations of this objective
\cite{MooreRab2007}.

\section{Discussion}

The complexity results derived herein are based on analysis of the kinematic contribution to optimal control gradient flows, and have neglected the dynamical contribution of the $\varepsilon(t) \rightarrow U(T)$ map to the scaling of search effort. For an arbitrary Hamiltonian, if one averages over many initial conditions and target gates/states, the relative convergence times of gate and state control optimization should correspond to those of the kinematic gradient flows \cite{RajWu2007}. However, since the Hamiltonian, and hence the properties of the $\varepsilon(t) \rightarrow U(T)$ map must change with system dimension, the absolute scaling for either problem may in principle deviate from that predicted here. The primary importance of the above results is that the \textit{relative} complexity scaling of these two problems will nonetheless be identical, since the contributions of the functional derivatives $\frac{\delta U(T)}{\delta \varepsilon(t)}$ to scaling, averaged over initial conditions and target gates, will be the same for each. As such, quantum pure state and gate control problems do belong to the same complexity class.

From a practical perspective, since the kinematic complexity of these problems is as low as possible, the scaling of control search effort should not prohibit the efficient application of gradient-based algorithms to high-dimensional systems. For state-to-state coherence transfer problems,
this prediction is borne out by both simulations and experimental evidence \cite{MooreRab2007}. However, the reported scaling of gate optimization simulations \cite{Kosloff2002} is worse than that predicted according to the present theory, suggesting that a reexamination of current quantum gate control algorithms is warranted. For control problems whose kinematic complexity class exceeds CLOG, it may be advantageous to apply matrix tracking algorithms (below) rather than gradient-based algorithms in order to achieve the above lower bounds. Evidence suggests that observable maximization problems starting from an initial nondegenerate mixed state belong to this category \cite{GregRab2007}.

For general $\rho_0$ and $\Theta$, the gradient flow for objective function $\Phi_1$ is an "isospectral" flow \cite{Faybu1991,Bloch1990, Bloch1995}. In N dimensions, this flow has N integrals of the motion that
are in involution, which is the classical definition of complete integrability for a dynamical system. From the point of view of the modern theory of integrable systems, the double bracket flow
can be shown to represent a type of Lax pair, a general form that can be adopted by all completely integrable dynamical systems \cite{Babelon2003}. As such, the system-independent complexity of all discrete quantum observable maximization problems can in principle be analytically determined \cite{RajWu2007}. In contrast, the kinematic gradient flows of the gate and observable control problems in classical mechanics \cite{WuRaj2007, Raj2007} are not integrable, and hence cannot be assigned to analytic complexity classes. 

In the present work, we have focused attention on the complexity of optimizations that employ local gradient search algorithms. The favorable scaling of the convergence times for kinematic gradient flows, and the scarcity of abnormal extremals in discrete quantum control problems \cite{Raj2007} motivate a rigorous definition of the universal complexity of quantum control problems in terms of the scaling of the expense of tracking such kinematic paths through elimination of the Hamiltonian-dependent matrix $\textmd{G}$ in the algorithmic step (\ref{vecu}). These global algorithms require an additional computational overhead of $N^4$ in order to invert the matrix $\textmd{G}$, as well as overhead for statistical estimation of the states or dynamical propagators based on experimental observations \cite{Raj2007}. Assignment of quantum control problems to complexity classes based on the convergence of such global algorithms, which involves extensive sampling over Hamiltonian space, is the subject of a separate work \cite{Raj2007b}.


\end{document}